\documentclass[twocolumn,showpacs,showkeys,superscriptaddress]{revtex4}
\usepackage{graphicx}
\usepackage{bm}
\usepackage{color}
\usepackage{amsmath}
\usepackage{amsfonts}
\usepackage{amssymb}
\usepackage{natbib}
\usepackage{latexsym}

\begin{document}

\title{Casimir force induced by electromagnetic wave polarization in
 Kerr, G\"odel and Bianchi--I spacetimes}

\author{Felipe A. Asenjo}
\email{felipe.asenjo@uai.cl}
\affiliation{Facultad de Ingenier\'ia y Ciencias,
Universidad Adolfo Ib\'a\~nez, Santiago 7491169, Chile.}
\author{Sergio A. Hojman}
\email{sergio.hojman@uai.cl}
\affiliation{Departamento de Ciencias, Facultad de Artes Liberales,
Universidad Adolfo Ib\'a\~nez, Santiago 7491169, Chile.}
\affiliation{Departamento de F\'{\i}sica, Facultad de Ciencias, Universidad de Chile,
Santiago 7800003, Chile.}
\affiliation{Centro de Recursos Educativos Avanzados,
CREA, Santiago 7500018, Chile.}


\begin{abstract}
Electromagnetic waves propagation on either rotating or anisotropic spacetime backgrounds (such as Kerr and G\"odel metrics, or  Bianchi--I metric) produce a reduction of the magnitude of Casimir forces between plates. These curved spacetimes behave as chiral or birefringent materials producing dispersion of electromagnetic waves, in such a way that right-- and left--circularly polarized light waves propagate with different phase velocities. Results are explicitly calculated for discussed cases. The difference on the wavevectors of the two polarized electromagnetic waves  produces an abatement of a Casimir force which depends on the interaction between the polarization of electromagnetic waves and the properties of the spacetime.
\end{abstract}


\maketitle

\section{Introduction}

The Casimir force, which acts on two (uncharged) conducting parallel plates in vacuum is, in general, attractive \cite{polder,casimir}. Casimir force can be explained as a consequence of quantum vacuum fluctuations, but it can be alternatively understood in terms of Van der Waals interactions \cite{jaffe}. Although this force is attractive in vacuum and depends only in the plates separation, a different behavior can be manifested under other conditions. When a medium made of different materials (or metamaterials),  with diverse properties or geometrical shapes, is inserted between the plates, a repulsive Casimir force may appear (see for instance, Refs.~\cite{lif,lif2,hoye,Kenneth,Zhao,lim,levin,rosa,Mathias,henkel,shao,alejo,munday,wil1,yaping}). The Casimir force can even produce a torque in the plates in optically anisotropic  materials \cite{somers}.

Recently, Jiang and Wilczek \cite{wil1} have shown that a tunable repulsive and attractive Casimir force can also be obtained when electromagnetic waves propagate in chiral material media between two conducting parallel plates. In such cases, the behavior of the Casimir force can be controlled by varying the electric permittivity and magnetic permeability of the material.
In this chiral material, right-- and left--circularly polarized electromagnetic waves propagate with wavevectors $k_+$ and $k_-$ respectively. In this material, both wavevectors are  described by $k_\pm=k\pm\delta k$, meaning that the difference  between the wavevectors of the two polarized waves is given by $k_+-k_-=2\delta k$. 
This difference is responsible for producing tunable  repulsive or attractive Casimir forces when the separation $l$ between plates is varied. 

In such cases, the Casimir energy is sensitive to this phase difference, and it turns out to be \cite{wil1}
\begin{eqnarray}
E_C=\int_0^\infty \frac{d\varpi}{2\pi}\int_{-\infty}^\infty \frac{d^2 k_\parallel}{4\pi^2} \ln \left[{1+e^{-4\varrho l}-2e^{-2\varrho l}\cos(2\delta k l)}\right]\, ,
\end{eqnarray}
where 
$\varpi=i\omega$, $\varrho=({\varpi^2+k_\parallel^2})^{1/2}$,  $k_\parallel=({k_1^2+k_2^2})^{1/2}$, with the wave frequency $\omega$, and the wavevector components $k_i$ ($i=1,2,3$). 
From here, the Casimir force per unit area ${\cal{F}}_C=- d E_c/dl$ experienced between two plates is
readily to be
\begin{eqnarray}\label{normalForceCas}
&&{{\cal{F}}_C}=\frac{1}{2\pi^3}\int_0^\infty d\varpi\int_{-\infty}^\infty d^2 k_\parallel \nonumber\\
&&\quad \left(\frac{\varrho e^{-4\varrho l}-\varrho e^{-2\varrho l}\cos(2\delta k l)-\delta k  e^{-2\varrho l}\sin(2\delta k l)}{1+e^{-4\varrho l}-2e^{-2\varrho l}\cos(2\delta k l)}\right)\, .
\end{eqnarray}
This force can be either attractive or repulsive depending on the magnitude of $2\delta k l$, and it has been shown in Ref.~\cite{wil1} that repulsive Casimir force may appear in Faraday and Optical active materials. On the contrary, in  a non-chiral material,  as there is no difference between  phase velocities of right-- and left--circularly polarized waves $\delta k=0$, the  force  \eqref{normalForceCas} becomes simply $-{\pi^2}/({240 l^4})$, recovering the value for vacuum Casimir force.

In Ref.~\cite{wil1}, only material media were considered. The main purpose of this work is  to show that gravitational fields described by stationary spacetimes (metrics with rotation, such as Kerr black holes or G\"odel universe), or by  anisotropic cosmological spacetimes (such as Bianchi--I metric),
produce a behavior similar to the one  appearing in chiral media when electromagnetic waves propagate on it.

Related gravitational effects due to spacetime curvature on Casimir force have been explored previously \cite{miltober,fuentes,elizal,artem,muniz,fulling,munaw,miltonpra,karim2}. Also, the effect of spacetime curvature correction due to Kerr metric on Casimir forces
has been studied for both scalar \cite{sorge,bezerra,nazari1,zhang,nouni} and electromagnetic fields \cite{nazari2}.
Our work  explores  how  the coupling of polarization with the angular momentum or anisotropy of spacetime  modifies the  Casimir force. We show this effects by  describing the analogy between electromagnetic wave propagation in curved spacetimes with the electromagnetic fields in media. We obtain solutions for Kerr, G\"odel and Bianchi--I spacetimes, where this coupling creates a difference between right-- and left--circularly polarized electromagnetic waves. We finally show how the chiral behavior of light on those spacetimes modifies the Casimir force.

\section{Maxwell equations in curved spacetime}

We start by describing the dynamics of electromagnetic fields on a gravitational background field (from now on we use natural units $c=\hbar=1$).
In general, covariant Maxwell equations in curved spacetime may be written as
\begin{equation}\label{equationMaxwVect0}
\nabla_\alpha F^{\alpha\beta}=0\, ,\quad \nabla_\alpha F^{*\alpha\beta}=0\, ,
\end{equation}
 in terms of  the antisymmetric electromagnetic field tensor $F^{\alpha\beta}$ and its dual $F^{*\alpha\beta}$. Here, $\nabla_\alpha$ is the covariant derivative defined by a metric $g_{\mu\nu}$. 

For the case of electromagnetic waves, several articles \cite{mas1,mas2,ase1,ase2,goshsen,skr,plebanski,chinmo} have shown that light does not propagate along null geodesics in curved spacetimes, giving rise to a dispersion relation that depends on their polarization. This occurs because gravitational fields behave as effective material media with non--trivial effective permeability and effective permittivity, both of them modifying the (vacuum) Maxwell equations. Thus, the electromagnetic field amplitude and polarization (besides its phase), couple to spacetime curveture.
This is explicitly shown by defining the corresponding electric 
$E_i=F_{i0}$, $D^i=\sqrt{-g} F^{0i}$, and
magnetic  $B^i=\varepsilon^{0ijk} F_{jk}$, $\varepsilon^{0ijk}H_k=\sqrt{-g} F^{ij}$ fields,
where $\varepsilon^{0ijk}$ is the Levi-Civita symbol and  $g$ the metric determinant with latin indices used to denote (three--dimensional) space coordinates.  The above vector fields are related by \cite{plebanski,mas1,ase1,goshsen,chinmo,gomb}
\begin{eqnarray}\label{Drepresentation}
D^i&=&\epsilon^{ij}E_j-\varepsilon^{0ijk}\mu_j H_k\, \nonumber\\
B^i&=&\epsilon^{ij}H_j+\varepsilon^{0ijk}\mu_j E_k\, ,
\end{eqnarray}
explicitly showing the analogy with electric and magnetic fields in the presence of a medium. Here 
\begin{equation}\label{permitivityper}
\epsilon^{ij}=-\sqrt{-g}\frac{g^{ij}}{g_{00}}\, , \quad {\mu_j}=-\frac{g_{0j}}{g_{00}}\, ,
\end{equation} 
are the effective  permittivity 
and the  effective vector permeability of the curved spacetime, respectively. Both are defined in terms of  spacetime metric (and its inverse $g^{\mu\nu}$). Using the above electric and magnetic fields, Maxwell equations \eqref{equationMaxwVect0} can now be written in a vectorial fashion as
\begin{eqnarray}\label{equationMaxwVect}
\partial_i D^i&=&0\, ,\quad \partial_0 D^i=\varepsilon^{0ijk}\partial_j H_k\, ,\nonumber\\
\partial_i B^i&=&0\, , \quad \partial_0 B^i=-\varepsilon^{0ijk}\partial_j E_k\, ,
\end{eqnarray}
where $\partial_0$ and $\partial_i$ stands for the time and spatial partial derivatives. 
Eqs.~\eqref{equationMaxwVect} are completly equivalent to the covariant Maxwell equations \eqref{equationMaxwVect0}.

An alternative and enlightening representation of electromagnetic fields in curved spacetime \cite{mas1,mas2} can be obtained by defining the (generalized) Riemann--Silberstein vectors 
\begin{eqnarray}
F_{j\pm}&=&E_j\pm i H_j\, ,\nonumber\\
S_\pm^j&=&D^j\pm i B^j\, ,
\end{eqnarray}
where the $+$ ($-$) symbol represents right (left) polarization of the electromagnetic fields. Using relations \eqref{Drepresentation}, we obtain that
\begin{equation}
S_{\pm}^j=  \epsilon^{ij} F_{j\pm}\pm i \varepsilon^{0ijk}\mu_jF_{k\pm}\, ,
\end{equation}
and therefore, Maxwell equations \eqref{equationMaxwVect} reduce to the simplest form 
\begin{eqnarray}
\partial_j S^j_\pm&=&0\, ,\nonumber\\
\pm i\, \partial_0 S_\pm^j&=&\varepsilon^{0jkm}\partial_k F_{m\pm}\, .
\end{eqnarray}
We can now use Riemann--Silberstein polarization vectors, to finally write Maxwell equations curved spacetime  as \cite{mas1}
\begin{equation}\label{GenMaxmasmenos}
\pm i\, \partial_0\left(\epsilon^{jk} F_{k\pm}\right)=\varepsilon^{0jkm}\left[\partial_k F_{m\pm}+\partial_0\left(\mu_k F_{m\pm}\right) \right]\, .
\end{equation}

Eqs.~\eqref{GenMaxmasmenos} are completely equivalent to Eqs.~\eqref{equationMaxwVect0}. From here is clear that, in general, both polarization states ($\pm$) propagate differently. Spacetime couples with different components of electromagnetic fields (this fact does not occur in vacuum), even producing rotation of the polarization state of light \cite{goshsen,ase1,ase2}. 
Isotropic and symmetric spacetimes (with $\mu_j=0$ and $\epsilon^{ij}=\epsilon\delta^{ij}$) allow null geodesic light propagation \cite{ase1,ase2}, while a general curved spacetime behaves  as an effective medium for travelling electromagnetic waves. Nevertheless, when $\mu_j\neq 0$ spacetime acts as a chiral material. Similarly, when $\epsilon^{ij}=\epsilon^i \delta^{ij}$, with $\epsilon^i\neq\epsilon^j$ for $i\neq j$, spacetimes  produce birefringence \cite{ase1,ase2}.

\section{Polarized electromagnetic waves in rotating and anisotropic spacetimes}
\label{Solution}

It can be straightforwardly shown that Kerr and G\"odel curved spacetimes, both having
$\mu_j\neq 0$,  produce different solutions for electromagnetic waves  with right-- or left--polarization. 
On the other hand,   cosmological anisotropy with   $\epsilon^{ij}\approx \epsilon^i\delta^{ij}$ and  $\mu_j=0$,
produces birefringence that also modifies the propagation  of polarized waves.

Thereby, these effects induce  changes in the magnitude of Casimir force, which are directly associated to the interaction of light polarization and  gravitational fields.

\subsection{Electromagnetic waves propagating in a slowly Kerr black hole background}

For the sake of simplicity,
we focus 
in the case of an electromagnetic wave travelling in the exterior of the gravitational field of a slowly rotating Kerr black hole 
with mass $M$ and spin $J$. This problem
was  solved by Mashhoon \cite{mas1}, showing that these propagating electromagnetic waves experience chiral effects in such a metric due to coupling with their polarization.

For a slowly rotating black hole, the stationary Kerr metric
in isotropic coordinates becomes \cite{mas1,misner}
\begin{eqnarray}\label{metricKerrslow}
ds^2&\approx&-\left(1-\frac{2M}{r}\right) dt^2-\frac{4J}{r^3}\left( x_1 dx_2-x_2dx_1\right) dt\nonumber\\
&&+\delta_{ij}dx^i dx^j\, ,
\end{eqnarray}
for a rotation in the $x_3$-direction, and with  $r=|{\bf x}|$. Its spin (angular momentum of magnitude $J$) is $J^j=\delta^{j3} J$ in these coordinates. 
 This intrinsic
angular momentum is measured far from the black hole \cite{misner}. 
Using metric \eqref{metricKerrslow} in \eqref{permitivityper}, 
we get that 
\begin{equation}\label{mediumemu}
\epsilon^{ij}\approx\delta^{ij}\left(1+\frac{2 M}{|{\bf x}|}\right)\approx\delta^{ij}\, ,\quad \mu_j\approx -2\varepsilon_{0jkm}\frac{J^k x^m}{|{\bf x}|^3}\, ,
\end{equation}
are the permittivity and permeability of
the effective medium for an electromagnetic wave propagating far from the black hole, with  $|{\bf x}|\sim D\gg M$, where $D$ is the given distance where the light propagates. For our slowly rotating metric, $|\mu|\sim J/D^2\ll 1$.
Therefore, Maxwell equations \eqref{GenMaxmasmenos}  reduce to \cite{mas1}
\begin{equation}\label{GenMaxmasmeno2}
\pm \omega\,  \delta^{jk}f_{k\pm}=\varepsilon^{0jkm}\left[\partial_k -i\, \omega\, \eta\,  \left(x_2 \delta_{1k}- x_1\delta_{2k}\right) \right]f_{m\pm} \, ,
\end{equation}
where we have used $F_{j\pm}(t, {\bf x})=f_{j\pm}({\bf x}) e^{-i\omega t}$, with  wave frequency $\omega$. Besides, $\eta=2 J/D^3$. Furthermore, the scale of the electromagnetic wave is much less than the one of the black hole, such that $\omega M\gg 1$.

 Solutions to Eq.~\eqref{GenMaxmasmeno2} represent right-- and left--circularly polarized electromagnetic waves propagating in the $x_2-x_3$ plane, very far from the Kerr black hole (where spacetime is almost flat). Performing the change of variables $f_{j\pm}({\bf x}) =\exp(i\omega \eta x_1 x_2)\Psi_{j\pm}({\bf x})$, from Eq.~\eqref{GenMaxmasmeno2}
we find
\begin{equation}\label{GenMaxmasmeno3}
\mp i \, \omega\,  \Psi^j_{\pm}=-i\, \varepsilon^{0jkm}\partial_j \Psi_{m\pm}+ 2\omega\,\eta\, x_1 \varepsilon^{0jkm} \delta_{2k} \Psi_{m\pm} \, .
\end{equation}

This equation can be readily solved by using a new change of variables
 $\Psi_{j\pm}({\bf x}) =\exp(i k_2 x_2+i k_3^\pm x_3)\phi_{j\pm}(x_1)$, where 
$k_{2}$ and $k_3^\pm$ are the  wavevectors in directions $x_2$ and $x_3$ respectively, for the corresponding polarizations. 
Each component of $\phi_{j\pm}$ satisfies different equations. The component along the $x_3$-direction has the solution $\phi_{3\pm}(\xi)\propto\exp(-\xi^2/2) H_{n^\pm }(\xi)$ \cite{mas1},
where $\xi=(2\eta \omega x_1+k_2)/\sqrt{2\eta\omega}$, and $H_{n^\pm}$ is the Hermite polynomial, with $n^\pm=0,1,2,...$. For this solution, the dispersion relation
\begin{equation}\label{dispere}
\omega^2-\left(k_3^\pm\right)^2\pm2\eta\, k_3^\pm=2\eta\,\omega(2n^\pm+1)\, ,
\end{equation}
must be satisfied. Solutions for the remaining components $\phi_{1\pm}$ and $\phi_{2\pm}$ are also known, and we refer the reader to Ref.~\cite{mas1} in order to find their thorough mathematical study.

Dispersion relation \eqref{dispere} defines the propagation of an electromagnetic wave in Kerr metric (and in a G\"odel universe as we will see). As $k_3^+\neq k_3^-$, this relation
shows that right-- and left--circularly polarized waves move with different phase velocities when propagating along the direction of the rotation axis of rotation of curved spacetime. 
In general, we find that the difference between the wavevectors is
\begin{eqnarray}\label{diffk3}
k_3^+-k_3^-&=& 2\delta k=2\eta+\sqrt{(\omega-\eta)^2-4\eta\omega n^+}\nonumber\\
&&\quad-\sqrt{(\omega-\eta)^2-4\eta\omega n^-}\, ,
\end{eqnarray}
which is a function of the frequency of the wave. However, because of our previous assumption on the nature of the black hole and the scales of electromagnetic waves,  we always find  that 
${\omega}/{\eta}={\omega M}{(M/D)^{-1}(J/D^2)^{-1}}\gg 1$,
describing waves in the high-frequency limit. In this case, the physical solutions for the electromagnetic wave components $\phi_{1\pm}$ and $\phi_{2\pm}$ imply that $n^+-1=n^-+1$ \cite{mas1}. All these allow us to find that the final wavelength difference becomes simply 
\begin{eqnarray}\label{diffk3b}
2\delta k&=& -2\eta\, .
\end{eqnarray}

Difference \eqref{diffk3b} is a physical consequence of the reaction of wave polarization to the rotation axis of the spacetime. Each spin of the circularly polarized wave couples differently to curvature. It is not an effect of coordinate choice, as the angular momentum of Kerr metric is well-defined as a measure of the frame dragging of this spacetime
\cite{misner,hartle}.
The above result shows that these left-- and right--circularly polarized electromagnetic waves propagate
in an fashion  analogous to waves experiencing Faraday rotation effect, which occurs in materials \cite{fara1,emunin} and magnetized plasmas \cite{kral}.

\subsection{Electromagnetic waves propagating in a slowly rotating G\"odel universe background}

Let us consider a G\"odel universe, with the stationary metric in cartesian coordinates \cite{mas1,ellishaw}
\begin{eqnarray}
ds^2&=&-dt^2+dx_1^2+dx_3^2\nonumber\\
&&+2\sqrt{2}\left[1-e^{\sqrt{2}x_1\Omega}\right]dx_2dt\nonumber\\
&&+\left[4e^{\sqrt{2}x_1\Omega}-e^{2\sqrt{2}x_1\Omega}-2 \right] dx_2^2\, ,
\end{eqnarray}
where  $\Omega$ is a constant related to the angular velocity of the rotating universe.

In the case of a slowly rotating universe, we have  $x_1\Omega \ll 1$, and from \eqref{permitivityper} we simply find
\begin{equation}
\epsilon^{ij}\approx \delta^{ij}\, , \quad {\mu_j}=-2\, x_1 \Omega\, \delta_{2j}\, ,
\end{equation} 
Using this in Eq.~\eqref{GenMaxmasmenos}, it is straightforward to  show that  electromagnetic waves propagates satisfying the following equation
\begin{equation}\label{GenMaxmasmeno4}
\mp i \, \omega\,  \Psi^j_{\pm}=-i\, \varepsilon^{0jkm}\partial_j \Psi_{m\pm}+ 2\omega\,\Omega\, x_1 \varepsilon^{0jkm} \delta_{2k} \Psi_{m\pm} \, ,
\end{equation}
with the transformation $F_{j\pm}(t, {\bf x})=\Psi_{j\pm}({\bf x}) e^{-i\omega t}$ and  wave frequency $\omega$. 

Equation \eqref{GenMaxmasmeno4}
is exactly the same than Eq.~\eqref{GenMaxmasmeno3} (replacing $\eta$ by  $\Omega$). Hence, we can find the same solutions, satisfying the dispersion relation \eqref{dispere}, and wavelength difference \eqref{diffk3b}.
Anew, the polarized behavior of light is a consequence of its coupling with the spacetime of G\"odel universe, whose rotation is a consequence of the total vorticity of the flow that generates it \cite{ellishaw}.

\subsection{Electromagnetic waves propagating in anisotropic cosmology}
 \label{anisotr}

Let us consider an anisotropic Bianchi--I cosmological model \cite{39}, with an interval given by 
\begin{equation}
ds^2=-dt^2+a_i^2 dx^i dx^i\, ,
\label{Bianmetric}
\end{equation}
in cartesian coordinates ($i=1,2,3$). Here, $a_i=a_i(t)$ are arbitrary time dependent functions, such as
$a_1\neq a_2\neq a_3$, in general. With this metric, we find from Eq.~\eqref{permitivityper} that $\mu_j=0$ and $\epsilon^{ij}=\epsilon^i\delta^{ij}$, with
\begin{equation}
\epsilon^i=\frac{a_1a_2a_3}{a_i^2}\, .
\end{equation}
This cosmology behaves in analogue fashion to a birefringent medium, as $\epsilon^1\neq\epsilon^2\neq \epsilon^3\neq \epsilon^1$ in general \cite{ase1}. 
Friedmann--Lema\^{i}tre--Robertson--Walker cosmology
is the particular case when $a_1=a_2=a_3=\epsilon_1=\epsilon_2=\epsilon_3$.

Using the general metric \eqref{Bianmetric}, Eqs.~\eqref{GenMaxmasmenos}
are now written  as
\begin{equation}\label{GenMaxmasmenosBianchi}
\pm \, \partial_0\left(\epsilon^i \delta^{ij} \zeta_{j\pm}\right)=\varepsilon^{0ijm}k_j \zeta_{m\pm}\, .
\end{equation}
where we have now considered that $F_{i\pm}(t,{\bf x})=\zeta_{j\pm}(t) e^{i{\bf k}\cdot{\bf x}}$, with constant ${\bf k}=(k_1,k_2,k_3)$. The three coupled Eqs.~\eqref{GenMaxmasmenosBianchi} can be solved for any component. As an example, let us obtain the wave equation for polarization $\zeta_{3\pm}$. First, manipulating \eqref{GenMaxmasmenosBianchi}, the following condition is  obtained
\begin{equation}\label{GenMaxmasmenosBianchicond}
\zeta_{2\pm}=\frac{-k_2k_3 \epsilon^3 \zeta_{3\pm}\pm k_1\partial_\xi(\epsilon^3\zeta_{3\pm})}{\epsilon^1 k_1^2+\epsilon^2 k_2^2}\, ,
\end{equation}
where the derivative $\partial_\xi$ is in terms of the effective cosmological time 
\begin{equation}\label{cosmotime}
\xi=\int\frac{dt}{\epsilon^1}\, .
\end{equation}
Using again \eqref{GenMaxmasmenosBianchicond} in Eqs.~\eqref{GenMaxmasmenosBianchi}, we finally can get
\begin{eqnarray}\label{waveEqBian}
&&\partial_\xi\left[\frac{\epsilon^1 k_1^2}{\epsilon^1 k_1^2+\epsilon^2 k_2^2}\partial_\xi  \left(\epsilon^3\zeta_{3\pm}\right)\right]\nonumber\\
&&\qquad-\partial_\xi^2\left(\epsilon^3\zeta_{3\pm}\right)\mp\partial_\xi\left(\frac{\epsilon^1 k_1k_2k_3}{\epsilon^1 k_1^2+\epsilon^2 k_2^2} \right)\epsilon^3\zeta_{3\pm}\nonumber\\
&&\qquad=\epsilon^1 k_2^2\left(1+ \frac{\epsilon^3 k_3^2}{\epsilon^1 k_1^2+\epsilon^2 k_2^2}\right)\zeta_{3\pm}\, .
\end{eqnarray}
From this equation can be readily seen that
different polarizations propagate at different speed when $k_1\neq0$, $k_2\neq 0$ and $k_3\neq 0$.
Similar equations to \eqref{waveEqBian}
can be found for $\zeta_{1\pm}$ and $\zeta_{2\pm}$. 

In order to obtain a sensible and simple solution, let us consider the case of an almost--isotropic cosmology, such that $a_2=a$, and 
$a_1=a_3=a(1+\vartheta/2)$, with $\vartheta\ll 1$. The anisotropy is in the direction of $k_2$.
 In this case, we get
$\epsilon^1=\epsilon^3=a$ and $\epsilon^2\approx a(1+\vartheta)$. Furthermore, let us assume for simplicity the particular case when $k_1=k_2=k_3=k$. Thus, Eq.~\eqref{waveEqBian} reduces to
\begin{eqnarray}\label{wavezeta32}
\partial_\xi^2\left(a\, \zeta_{3\pm}\right)+\frac{1}{2}\partial_\xi\vartheta\, \partial_\xi\left(a\, \zeta_{3\pm}\right) +\lambda^\pm a\,  \zeta_{3\pm}=0\, ,
\end{eqnarray}
where
\begin{equation}
\lambda^\pm=\left(3-\frac{\vartheta}{2}\right)k^2\mp \frac{k}{2}\partial_\xi\vartheta\, .
\end{equation}
The different behavior of polarizations is due to only the variations of the anisotropy.
Eq.~\eqref{wavezeta32} can be  solved for $\zeta_{3\pm}$, as \cite{birrel}
\begin{equation}\label{solmasgenecosmo}
\zeta_{3\pm}(t)=\frac{\zeta_{3o}}{a}\exp\left(-\frac{\vartheta}{4}\right) \exp\left(-i\int \frac{W_{3}^{\pm}}{a}{dt}\right)\, ,
\end{equation}
where $\zeta_{3o}$ is a constant, and $W_{3}^{\pm}$ fulfill the equations
\begin{eqnarray}\label{disperanisotr}
\left(W_{3}^{\pm}\right)^2&=&\frac{3}{4}\left(\frac{\partial_\xi W_{3}^{\pm}}{W_{3}^{\pm}}\right)^2-\frac{\partial_\xi^2 W_{3}^{\pm}}{2\, W_{3}^{\pm}}\nonumber\\
&&+\lambda^\pm-\frac{(\partial_\xi\vartheta)^2}{8} -\frac{\partial_\xi^2\vartheta}{4}\, .
\end{eqnarray}
Similar solutions can be found for $\zeta_{1\pm}$ and $\zeta_{2\pm}$. 

The behavior of $\zeta_{3\pm}$
depends on $W_{3}^{\pm}$ by its evolution given in 
Eq.~\eqref{disperanisotr}. This equation
can be considered the dispersion relation for that particular propagation in this anisotropic cosmology. Hence,  the wave frequency for each polarization
can be defined as $\omega_{3}^{\pm}= W_{3}^{\pm}/a$. This definition takes into account the cosmological redshift.
 An approximated solution can be obtained assuming that variations of $W_{3}^{\pm}$ are negligible compared with variations of $\vartheta$. In this case, we get
\begin{equation}
a\, \omega_{3}^{\pm}\approx \sqrt{\lambda^\pm}\approx \sqrt{3k^2\mp \frac{k}{2}\partial_\xi\vartheta}\, .
\end{equation}
Wave frequency can be written as $a\, \omega_{3}^{\pm}=\sqrt{3} k\pm \delta k$, as a deviation from a light--like behavior when $k\gg\partial_\xi\vartheta$. Thereby
\begin{eqnarray}\label{diffanisotropic}
a\left(\omega_{3+}-\omega_{3-}\right)&=&2\delta k\nonumber\\
&=& \sqrt{3k^2- \frac{k}{2}\partial_\xi\vartheta}-\sqrt{3k^2+ \frac{k}{2}\partial_\xi\vartheta}\nonumber\\
&\approx& -\frac{a\, \partial_0\vartheta}{2\sqrt{3}}\, .
\end{eqnarray}
Under the above assumption, it is clear that different polarizations propagate differently only by their interaction with the anisotropic characteristic of this cosmology.

\section{Casimir force due  polarized electromagnetic waves}
\label{Repulsivesucs}

In Ref.~\cite{wil1} it was shown that polarized electromagnetic waves can give origin to repulsive or attractive Casimir forces when the wavevector of their left and right polarizations fulfill $k^+\neq k^-$ in chiral media. This is exactly the behavior of differences  \eqref{diffk3b} and  \eqref{diffanisotropic} 
for polarized electromagnetic waves propagating in a Kerr and G\"odel  background metrics and in anisotropic cosmologies, respectively. 

As a consequence, we can treat these curved spacetimes as analogue media for photon propagation with the sole purpose of calculating the associated Casimir force \cite{nouni,nazari2}. In our case, polarized high--frequency electromagnetic waves fulfilling differences \eqref{diffk3b} and \eqref{diffanisotropic}, modify the Casimir force decreasing its intensity, so it becomes less attractive.

\subsection{Casimir force due to a rotating spacetime}

In this case, the wavelenght difference is $\delta k=-\eta$ for Kerr sacetimes, and $\delta k=-\Omega$ for G\"odel spacetimes. Now, let us consider two parallel plates with a length separation $l$, at some distance $D\gg l$ far from the rotating black hole. Plates size are negligible compared to black hole dimensions.
The plates are perpendicular to the rotation axis of the rotating spacetime, and in between them, polarized electromagnetic waves propagate. 
The Casimir force per unit area \eqref{normalForceCas}
experienced between two  plates is then
\begin{eqnarray}\label{normalForceCasK}
&&{{\cal{F}}_C}=\frac{1}{2\pi^3}\int_0^\infty d\varpi\int_{-\infty}^\infty d^2 k_\parallel \nonumber\\
&& \left(\frac{\varrho e^{-4\varrho l}-\varrho e^{-2\varrho l}\cos(2\eta l)-\eta e^{-2\varrho l}\sin(2\eta l)}{1+e^{-4\varrho l}-2e^{-2\varrho l}\cos(2\eta l)}\right)\, ,
\end{eqnarray}
where we have used $\eta$ for both Kerr and G\"odel cases.
Although this result is general, we must notice that for Kerr metric
$\eta l=\left({2J}/{D^2}\right)\left({l}/{D}\right)\ll1$,
whereas for a G\"odel universe 
$\Omega l=(D\Omega)\left({l}/{D}\right)\ll1$.
Using this, we can explicitly evaluate \eqref{normalForceCas} to obtain 
\begin{eqnarray}\label{normalForceCas2}
{\cal{F}}_C\approx -\left(1-\frac{10}{\pi^2}\eta^2 l^2\right)\frac{\pi^2}{240 l^4}\, ,
\end{eqnarray}
Casimir force per unit area  \eqref{normalForceCas2} shows that spacetimes  describing rotating bodies (such as a Kerr black hole and G\"odel universe) induce a reduction on the magnitude of  attractive Casimir forces. This effect is due to  the chiral behavior of
polarized electromagnetic  waves triggered by the coupling of different circularly polarized waves and the rotation of spacetime, in an analogue fashion to what occurs in a chiral medium. 

Strictly speaking, in order to obtain the result \eqref{normalForceCas2}
we have used the flat--spacetime calculation for electromagnetic waves in a medium, developed in Ref.~\cite{wil1}. However, forces \eqref{normalForceCasK} and \eqref{normalForceCas2} are evaluated on curved spacetimes. This is possible to  be done because we have taken into consideration the effect of spacetime curvature into the electromagnetic wave propagation, by defining the effective medium described by the permittivity and permeability  \eqref{mediumemu}. Therefore, a medium treatment for Casimir energy becomes completely analogous to waves in curved spacetime. On the other hand, Casimir effect has been already studied in this limit on Kerr spacetimes for scalar fields \cite{sorge,sorge2,nazari1,bezerra}, and electromagnetic fields \cite{nazari1}. For both kind of fields, the gravitational correction to Casimir energy depends on $\mathcal{O}(M/D)$, producing a decreasing of the attractive force. The above used assumptions in our calculations require that $M/D\ll 1$, as it can be seen in \eqref{mediumemu}. This implies that we are neglecting the gravitational potential (Schwarzschild--like)  correction to the Casimir force in our calculations. However, the net effect of the rotation spacetime does not vanish, as $\eta$ is independent of the $M$. Previous results for the Casimir force produced by electromagnetic fields in a weak Kerr gravitational field have shown the $M/D$ dependence 
without considering the polarization coupling with the rotation of the black hole  \cite{nazari1}. As soon as this effect is taken into account, new corrections of order $\eta^2l^2$, not previously envisaged, emerge as a source of reduction of the magnitude of this force.

\subsection{Casimir force due to anisotropic cosmology}

In this case, the 
difference between left-- and right--polarization is produced by anisotropies in spacetime, and is equal to $\delta k=-a \partial_0\vartheta/(4\sqrt{3})$.
 This induces a modification on the Casimir force \eqref{normalForceCas} for two parallel plates with a proper length separation $a\, l$. 

Consider a separation smaller than the scale of variation of the anisotropy, $a\, l\, \partial_\xi \vartheta\ll 1$. 
In this case,  force \eqref{normalForceCas} for the $\zeta_{3\pm}$ polarizations
 is simplified to 
\begin{eqnarray}\label{normalForceCasCos}
{\cal{F}}_C\approx -\left(1-\frac{5 a^4}{24\pi^2}\left(l\, {\partial_0\vartheta}\right)^2\right)\frac{\pi^2}{240 \, a^4 l^4}\, ,
\end{eqnarray}
showing how anisotropic cosmology variations reduces the strength of Casimir force due to electromagnetic interactions, as this expanding spacetime behaves as a medium.
In the case of an isotropic cosmology with $\vartheta=0$, Casimir force \eqref{normalForceCasCos}
reduces to its known $\propto a^{-4}$ behavior \cite{romeroVB}.

\section{Discussion}

Results described in this work belong to a family of different behaviors of
 Casimir forces due to spacetime curvature. These have been studies in de Sitter \cite{elizal}, Schwarzschild \cite{fuentes,muniz,munaw}, G\"odel \cite{shojai}, wormhole \cite{artem} spacetimes, in cosmology \cite{romeroVB,romeroVB2,szy},
or in quantum cosmology \cite{miltober},
 due to  scalar \cite{sorge,sorge2,bezerra} and vectorial fields \cite{nazari1},
 also showing that Casimir energy complies with the equivalence principle \cite{fulling,miltonpra,karim2}. 

However, and differently, in this work we show that polarization of electromagnetic vacuum fluctuating fields interacting with curved spacetime (in a similar fashion to what occurs in the presence of materials)  also modify the Casimir force. 
This effect manifests itself (in the calculations) only when electromagnetic waves are not treated as light rays (in the eikonal limit).
It is the coupling between the spacetime curvature with the electromagnetic wave extended properties which modify the  strength of Casimir forces, in addition to the curvature effects produced by the gravitational field. In this way, the spacetime background  has a direct physical consequence that cannot be replicated by  scalar fields or electromagnetic fields in the eikonal limit.

Finally, forces \eqref{normalForceCas2} and \eqref{normalForceCasCos} hint that a more general spectrum of Casimir forces can be found for propagating electromagnetic waves in these kinds of curved backgrounds with inequivalent directions.
The results presented in this article can be generalized to any spacetime with $\mu_j\neq 0$ or non--constant $\epsilon^{jk}$. In general, for high--frequency electromagnetic waves in rotating spacetimes, the difference between the propagation of different polarizations should be proportional to $|\mu|$, or variations of $|\epsilon|$. Those effects create the small corrections to  the Casimir force. However,  one should expect to have larger effects if more general solutions of wave equation \eqref{GenMaxmasmenos} can be found. Whenever the current assumptions can be relaxed, the effect on different propagation for different polarizations of electromagnetic waves with longer wavelengths could be solved. Then it would be possible to study the repulsive effect on Casimir forces due to curved spacetimes, in an analogue fashion to what occurs in general chiral materials \cite{wil1}.
 These endeavors are left as the subject for future research.

\begin{acknowledgments}

F.A.A. thanks Fondecyt-Chile Grant No. 1180139. 
\end{acknowledgments}

\end{document}